\documentclass[
 aip,
jcp,
 amsmath,amssymb,
 reprint,%
 longbibliography
 ]{revtex4-1}

\usepackage{graphicx}
\usepackage{dcolumn}
\usepackage{bm}
\usepackage[colorlinks=true,linkcolor=blue,urlcolor=blue,citecolor=blue]{hyperref} 

\usepackage[utf8]{inputenc}
\usepackage[T1]{fontenc}
\usepackage{mathptmx}
\usepackage{etoolbox}
\usepackage{color}
\usepackage[normalem]{ulem}

\newcommand{\del}[1]{\sout{#1}}  
\renewcommand{\del}[1]{}  

\newcommand{\Cite}[1]{Ref.~\onlinecite{#1}}
\newcommand{\fig}[1]{Fig.~\ref{#1}}

\newcommand{\Fig}[1]{Figure~\ref{#1}}
\newcommand{\eq}[1]{Eq.~(\ref{#1})}

\newcommand{\roundbk}[1]{\left({#1}\right)}
\newcommand{\squarebk}[1]{\left[{#1}\right]}

\newcommand{\dE}{\Delta E}

\makeatletter
\def\@email#1#2{%
 \endgroup
 \patchcmd{\titleblock@produce}
  {\frontmatter@RRAPformat}
  {\frontmatter@RRAPformat{\produce@RRAP{*#1\href{mailto:#2}{#2}}}\frontmatter@RRAPformat}
  {}{}
}%
\makeatother

\begin{document}

\preprint{AIP/123-QED}


\title[Penetration of surface effects on structural relaxation and particle hops in glassy films]{Penetration of surface effects on structural relaxation and particle hops in glassy films}
\author{Qiang Zhai} 
\affiliation{MOE Key Laboratory for Nonequilibrium Synthesis and Modulation of Condensed Matter,  School of Physics, Xi'an Jiaotong University, Xi'an, Shaanxi, 710049, China}
\author{Hai-Yao Deng}
\affiliation{School of Physics and Astronomy, Cardiff University, 5 The Parade, Cardiff CF24 3AA, Wales}
\author{Xin-Yuan Gao}
\affiliation{Department of Physics, The Chinese University of Hong Kong, Shatin, New Territories, Hong Kong, China}
\author{Leo S.I. Lam}
\affiliation{Department of Mechanical Engineering, Hong Kong Polytechnic University, Hong Kong, China}
\author{Sen Yang}
\altaffiliation{ yangsen@mail.xjtu.edu.cn}
\affiliation{MOE Key Laboratory for Nonequilibrium Synthesis and Modulation of Condensed Matter,  School of Physics, Xi'an Jiaotong University, Xi'an, Shaanxi, 710049, China}

\author{Ke Yan}
\altaffiliation{ yanke@mail.xjtu.edu.cn}
\affiliation{School of Mechanical Engineering, Xi'an Jiaotong University, Xi'an, Shaanxi,710049,China}
\author{Chi-Hang Lam}
\altaffiliation{C.H.Lam@polyu.edu.hk}
\affiliation{Department of Applied Physics, Hong Kong Polytechnic University, Hong Kong, China}

\date{\today}

\begin{abstract}
  A free surface induces enhanced dynamics in glass formers.
We study the dynamical enhancement of glassy films with a distinguishable-particle lattice model of glass free of elastic effects. We demonstrate that the thickness of the surface mobile layer depends on temperature differently under different definitions, although all are based on local structure relaxation rate. The rate can be fitted to 
  a double exponential form with an exponential-of-power-law tail. Our approach and results exclude elasticity as the unique mechanism for the tail.
  Layer-resolved particle hopping rate, potentially a key measure for activated hopping, is also studied but it exhibits much shallower surface effects.
\end{abstract}

\maketitle

\section{Introduction}
It is well known that a glass former can retain a liquid-like free surface even though its bulk may well have vitrified ~\cite{baschnagel2005,ediger2013review,napolitano2017review,mckenna2017review,roth2021review,rodriguez2022review,cangialosi2024review}.
Of films with free surfaces, a reduction in  the glass transition temperature $T_\mathrm{g}$ accompanied by accelerated structural relaxation near the surface region has been observed \cite{keddie1994,fakhraai2008,yang2010,zhu2011,forrest2014}. 
Such surface-enhanced dynamics is confined to a surface mobile layer, over which local structural relaxation is accelerated. Its thickness has been studied in experiments \cite{ilton2009,paeng2011,yuan2022} and simulations \cite{varnik2002pre,hanakata2012,lang2013,lam2013,shavit2014,lam2018film}. However,  estimated values of the thickness and its  temperature dependence can vary significantly among different definitions and are  not yet fully understood. 

A deeper question concerns the mechanism of the enhanced dynamics and the spatial functional form of its penetration into the interior of a film. It was argued that the local structural relaxation time, i.e. $\alpha$-relaxation time, $\tau_\mathrm{\alpha}(z)$ can shed light on the general question of the glass transition ~\cite{long2001,ellison2003,pye2010,salez2015,phan2019,white2021,ghan2023,herrero2024}. Here $z$ is the depth from the free surface of a film.  Yet, direct experimental tests of local properties against theoretical predictions are technically difficult, given the limited spatial resolution~\cite{ellison2003,baglay2017b}. Recently, accurate measurements of $\tau_\alpha(z)$ from molecular dynamics (MD) simulations are reported to support a double-exponential form with an exponential-of-power-law tail given by
\cite{ghan2023}
\begin{equation}
  \tau^{-1}_\alpha(z) = (\tau_\alpha^\mathrm{bulk})^{-1}
   \exp \squarebk{ a\exp(-z/\xi_0)+b/z}, \label{dex0}
\end{equation}
more often written as
\begin{equation}
  \ln(\tau_\alpha^\mathrm{bulk}/\tau_\alpha(z)) = a\exp(-z/\xi_0)+b/z, \label{dex}
\end{equation}
where $\xi_0$ measures the mobile layer thickness, $\tau_\alpha^\mathrm{bulk}$ is the bulk $\alpha$-relaxation time, and $a$ and $b$ are constants. Based on an elastically
cooperative nonlinear Langevin equation (ECNLE) theory \cite{phan2019}, the double-exponential term, found also in early MD simulations~\cite{scheidler2003}, can be derived from an empirical layer-by-layer facilitation-like process for surface effects, while the tail can follow from a local softening due to an elastic coupling in three dimensions to the free surface \cite{ghan2023}. Although the power-law tail covers only half a decade of scaling and may require further scrutiny, the intriguing two-component form seems evident from the accurate MD results.
The functional form of \eq{dex} was claimed as a unique signature for the relevance of elastic field in glass \cite{ghan2023}. This echoes a recent increase of interest  \cite{tahaei2023,hasyim2024} in  the possible dominating role of elasticity in glassy dynamics \cite{dyre2006review,schweizer2019}.

Lattice models of glass in general enjoy superior computational efficiency \cite{garrahan2011review,ritort2003review}.
Here we use a distinguishable particle lattice model (DPLM) \cite{zhang2017,lee2020} to study the relaxation dynamics of supported glassy films. {The DPLM is fundamentally a free-volume model. It generalize multi-species model for glass. Emergent facilitation is seen without explicit facilitation rules.} The DPLM has been able to reproduce a variety of characteristic phenomena for bulk samples of glass, including slow relaxation \cite{zhang2017}, Kovacs paradox~\cite{lulli2020} and effect~\cite{lulli2021}, kinetic and thermodynamic fragility~\cite{lee2020}, heat capacity overshoot \cite{lee2021}, two-level systems \cite{gao2022}, diffusion coefficient power-laws \cite{gopinath2022}, and Kauzmann's paradox \cite{gao2023}. The same model has also  successfully simulated surface enhanced mobility \cite{zhai2024} and heat capacity  \cite{zhai2025} of glassy films.
The importance of adopting an extensively tested lattice model is to ensure compatibility among explanations of a wide range of glassy phenomena. Mechanisms considered in this study based on the DPLM are thus also consistent with the aforementioned list of characteristics of glass in both bulk and film geometries.  

In this paper, we measure and contrast estimates of the thickness of the surface mobile layer from DPLM simulations based on several commonly used definitions.
We demonstrate that the local $\alpha$-relaxation time $\tau_\alpha(z)$ measured for DPLM films can as well be fitted by Eq.~(\ref{dex}), despite the absence of elastic effects. 
Finally, we show that the particle hopping rate from the DPLM admits much weaker surface enhancement than the relaxation time and the implications will be discussed.

\section{Model} \label{sec2}

\begin{figure}[tb]
 \begin{centering}
  \includegraphics[width=1\columnwidth]{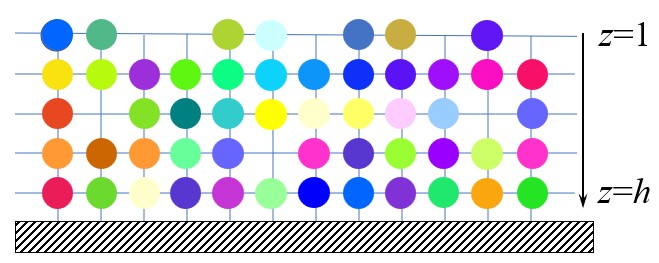}
\caption{A schematic drawing of the lattice model of supported film.}
\label{fig:model}
\end{centering}
\end{figure}

We follow the construction of DPLM films and all model parameters introduced previously for mobility and heat capacity measurements \cite{zhai2024,zhai2025}. Specifically, $N$ distinguishable particles of each of their own type  populate a square lattice of ribbon geometry, with thickness $h$ and length $L=1000$, in two dimensions (2D). Each lattice site can be occupied by either a particle or a void (non-occupied), as shown in \fig{fig:model}. We apply periodic boundary conditions along the direction of $L$ and fixed bounding walls along the direction of $h$. We designate $z=1$ for the particles in contact with the vacuum represented by the upper wall and $z=h$ for the particles that {sit on top of the substrate modeled by the lower wall}. 
The energy of the model is given by
\begin{equation}
E=\sum_{<i,j>}V_{s_is_j}n_in_j+\epsilon_\mathrm{bot}\sum_{i: z_i = h}n_i+\epsilon_\mathrm{top}\sum_{i: z_i = 1}n_i~. \label{E}
\end{equation}
Here, the particle type at the site $i$ is labeled $s_i$. The interaction energy between the particle at the site $i$ and the particle at the adjacent site $j$ is given by $V_{s_is_j}$. The occupation number $n_i=1$ at site $i$ if site $i$ is occupied by a particle, otherwise $n_i = 0$ if the site is empty. The $z$-coordinate (i.e. depth) of site $i$ is given by $z_i$. We use $\epsilon_\mathrm{top}=1.124$ to account for the excess interfacial energy experienced by particles in $z=1$ and $\epsilon_\mathrm{bot}=-0.5$ for particles in $z=h$, respectively. With the parameters chosen above, it has been shown \cite{zhai2024} that the particle density at $z > 1 $ equilibrates at a constant value of $0.99$, and drops sharply to $0.2$ at $z=1$. {The substrate is then dynamically neutral approximately at the studied temperatures. Notice that the first term of \eq{E} resembles the definition of a lattice gas model at first glance. The crucial difference is that the pair-interaction energy $V_{s_is_j}$ takes a random value for particle types $s_i$ and $s_j$ sampled at the beginning of a simulation. This definition introduces a disorder quenched in the configuration space so that the same interaction energy always applies for any given local configuration. This  ultimately leads to emergent glassy phenomena.} We sample $V_{s_is_j}$ from a \emph{a priori} distribution, 
\begin{equation}\label{eq:gv}
g(V)=G_0/\roundbk{V_1-V_0}+(1-G_0)\delta(V-V_1),
\end{equation} 
where $V_0=-0.5$, $V_1=0.5$ and $G_0=0.7$ are parameters chosen for the study \cite{zhai2024}. {The value of $G_0$ in \eq{eq:gv} has been demonstrated to control the fragility of glass formers~\cite{lee2020}.}

We simulate void-induced equilibrium dynamics with the Metropolis algorithm. The rate for a particle hopping to an adjacent void is given by
\begin{equation}
w=w_{0} \exp \squarebk{-\dE \Theta(\dE)/k_BT},
\end{equation}
where $\dE$ is the energy change due to the hop, { $\Theta(\dE)=1$ if $\dE>0$ or $\Theta(\dE)=0$ otherwise.} We set the attempt frequency $w_{0} = 10^{6}$. We use $k_B=1$, so the temperature has the same dimension as energy. The adopted algorithm guarantees detailed balance. 

We use the swap algorithm (swapping among all particles and voids) to accelerate the equilibration process \cite{ninarello2017,gopinath2022} . More than $10^{5}$ swap attempts per site are performed to equilibrate the energy and depth-dependent particle density. Before taking measurements, another $10^{5}$ swap attempts per site are performed to confirm that equilibrium has been attained. 

\section{Relaxation rate and mobile layer thickness}

\begin{figure}[tb]
 \begin{centering}
  \includegraphics[width=1\columnwidth]{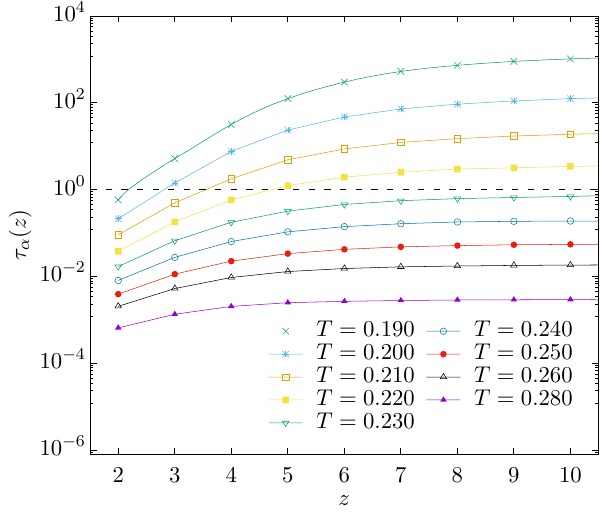}
\caption{Structural relaxation time, $\tau_\alpha(z)$, for a film of thickness $h=45$ at various values of temperature $T$. The free surface is {adjacent to} particles at $z=1$. The dashed line sets an empirical reference observation time $\tau_0=1$.}
\label{fig:talpha}
\end{centering}
\end{figure}

\begin{figure}[tb]
 \begin{centering}
  \includegraphics[width=1\columnwidth]{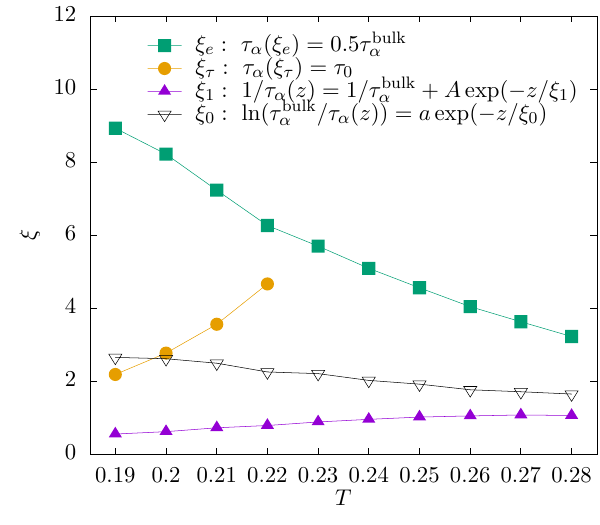}
  \caption{Thicknesses of surface mobile layer based in different definitions against temperature $T$ for a film of $h=45$.
  }
\label{fig:ht}
\end{centering}
\end{figure}

We measure from DPLM simulations the local overlap function $q(\Delta t,z)$, which gives the probability that a particle in layer $z$ is at the original position after a duration $\Delta t$. In \fig{fig:talpha}, we display the layer-resolved relaxation time $\tau_\alpha(z)$ defined via $q(\Delta t,z) = 1/e$ at $\Delta t=\tau_\alpha(z)$ \cite{zhai2024}.  
As seen, a slope of $\tau_\alpha(z)$ extends towards the bulk and increases as $T$ is reduced from $0.28$ to $0.19$. This is consistent with MD simulations \cite{varnik2002pre} and signifies the penetration of fast surface dynamics into the interior of the film.

Using $\tau_\alpha(z)$ from \fig{fig:talpha}, we calculate the mobile layer thickness $\xi_0$ from the double-exponential term in \eq{dex}. Alternatively, 
we define a different mobile layer thickness $\xi_{\tau}$ according to {\cite{paeng2011,lang2013,white2021}},

\begin{equation}\label{htt}
\tau_\alpha(\xi_{\tau})= {\tau_0},  
\end{equation}
based on an empirical reference relaxation time $\tau_0$. We have used $\tau_0=1$. {Beyond the length scale set by $\xi_\tau$, the dynamics is considered frozen for a practical observation time $\tau_0$. It is thus often more relevant in interpreting experimental results \cite{paeng2011}.}
In addition, one may introduce a third definition of thickness $\xi_e$ according to \cite{hanakata2012,lang2013,shavit2014}
\begin{equation}\label{htT}
\tau_\alpha(\xi_e)=0.5 \tau_\alpha^\text{bulk},
\end{equation}
where the mobility enhancement is gauged by the empirical prefactor 0.5 for the reduction of the {relaxation time} from the bulk value $\tau_\alpha^\text{bulk}$. {Such a definition then quantifies the width with dynamics significantly faster than the bulk dynamics.}
Finally, from the gradient of $\tau_\alpha(z)$ close to the free surface, we define a width $\xi_1$ by a single-exponential form \cite{lam2018film} 
\begin{equation}\label{hte}
  \tau_\alpha(z)^{-1}=(\tau_\alpha^\text{bulk})^{-1}+Ae^{-z/\xi_1}.
\end{equation}
{Both $\xi_0$ and $\xi_1$ quantify the characteristic decay lengths in the convergence of depth-resolved relaxation rate to its bulk value, according to different functional forms assumed by \eq{dex} and \eq{hte} respectively.}

\Fig{fig:ht} plots DPLM results on the thicknesses described above.
As temperature decreases, we observe that $\xi_\tau$ and $\xi_1$ decrease while $\xi_0$ and $\xi_e$ increase.
The temperature dependence of $\xi_\tau$, $\xi_0$, and $\xi_e$ has been studied before. 
Specifically,  experiments  on $\xi_\tau$ \cite{paeng2011} and MD results on $\xi_\tau$ \cite{lang2013}, $\xi_0$ \cite{scheidler2003} and $\xi_e$
\cite{hanakata2012,lang2013,shavit2014} have all reported trends in good agreement with those from the DPLM in 
\fig{fig:ht}.

\begin{figure}[!t]
 \begin{centering}
  \includegraphics[width=1\columnwidth]{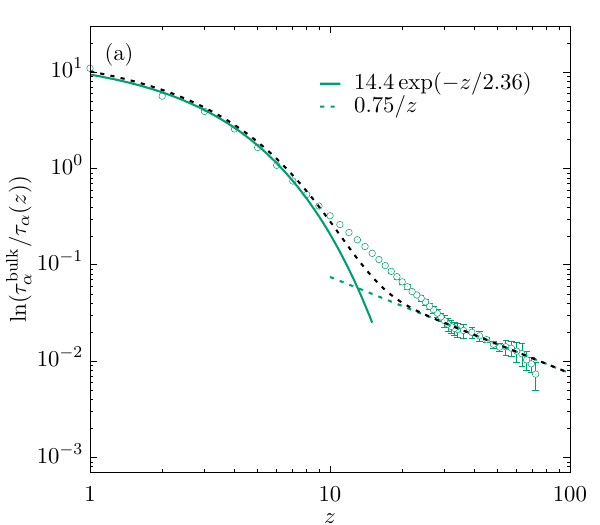}
  \includegraphics[width=1\columnwidth]{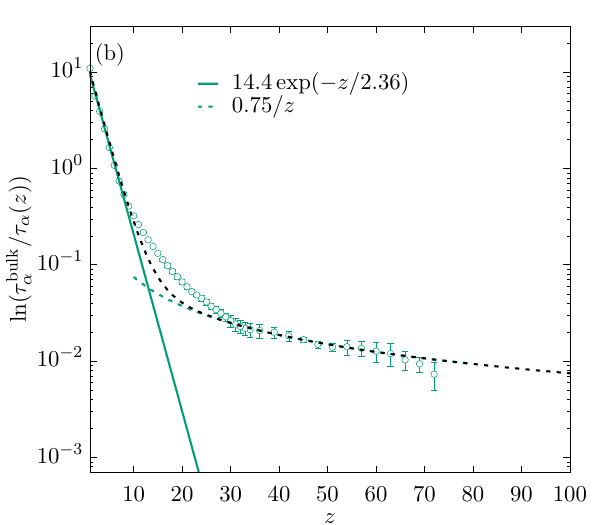}
  \caption{Plot of $\ln(\tau_\alpha^\mathrm{bulk}/\tau_\alpha(z))$
    versus depth $z$ in (a) log-log and (b) semi-log scales for a film of thickness $h=100$ at $T=0.21$. {Two asymptotic fits to the data are also displayed. The black line indicates the sum of the two fits.} }
\label{fig:de}
\end{centering}
\end{figure}

{We have measured surface mobile layer thicknesses $\xi_0$, $\xi_\tau$, $\xi_e$, and $\xi_1$ defined in Eqs. (\ref{dex}), (\ref{htt}), (\ref{htT}), and (\ref{hte}) respectively.
Our results show that $\xi_\tau$ and $\xi_1$ increase with temperature, while $\xi_0$ and $\xi_e$ decrease with temperature.} Such opposite trends are consistent with previous studies as explained above and have been discussed based on Adam-Gibbs theory \cite{lang2013} and cooperative free volume theory \cite{white2021}.
The reproduction of these trends here further supports the applicability of the DPLM on {surface-enhanced} dynamics in glassy films \cite{zhai2024,zhai2025}, in addition to several other glassy phenomena demonstrated previously \cite{lee2020,lulli2020,lulli2021,lee2021,gao2022,gopinath2022,gao2023}.

The different properties of the various thicknesses demonstrate that a full characterization of surface enhanced mobility in general requires the whole function of layer-resolved relaxation time $\tau_\alpha(z)$, while an individual thickness reflects only certain features of the dynamics.  In particular, $\xi_e$ captures the length scale over which the system cannot maintain its bulk dynamics.  It may be most closely related to the dynamic correlation length of glass in bulk, which is expected to increase as temperature decreases. In contrast, $\xi_1$ is a decay length of enhanced dynamics close to the surface. It dictates the length scale of the layer dominating surface flow and is thus most relevant for quantifying surface transport \cite{yang2010,lam2013}. Its value decreases as temperature decreases, signifying that transport is dominated by an increasingly thinner layer.

To further analyze the functional form of $\tau_{\alpha}(z)$, we have measured $\tau_{\alpha}(z)$ from supported DPLM films of thickness $h=100$. We determine $\tau_\alpha^\text{bulk}$ accurately from simulations of bulk samples so that statistical errors are dominated by those of $\tau_{\alpha}(z)$. \Fig{fig:de} plots $\log(\tau_\alpha^\mathrm{bulk}/\tau_\alpha(z))$ against $z$. The result strikingly resembles those from recent MD measurements \cite{ghan2023} and can be reasonably fitted to \eq{dex} with a double-exponential form in the short range and an upward-turning tail at $z \agt {30}$.

\section{Weak surface effects on particle hops}
\begin{figure}[!tb]
 \begin{centering}
  \includegraphics[width=1\columnwidth]{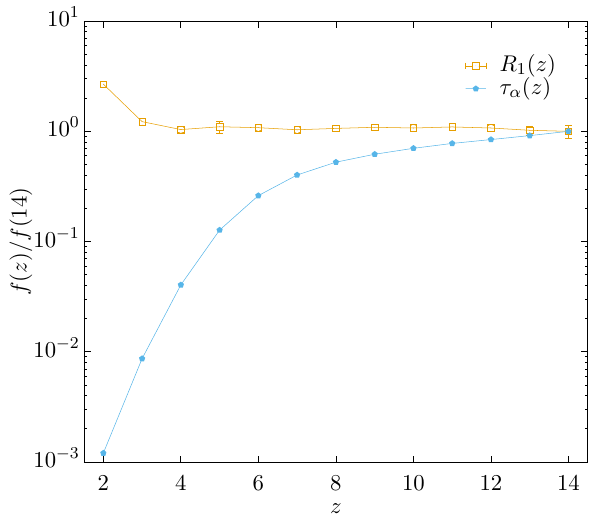}
\caption{{Particle hopping rate $R_1(z)$ against depth $z$ for a film of thickness $h=15$ at temperature $T=0.20$. Also plotted is structural relaxation time $\tau_\alpha(z)$} under the same conditions. The data is normalized by the respective quantity at $z=14$.} 
\label{fig:hop1}
\end{centering}
\end{figure}
Despite strong surface effects on the relaxation time, we next show that they are much weaker on particle hopping rate. At deep supercooling, particle dynamics in glass formers are dominated by activated particle hops as revealed by the emergence of secondary and higher-order peaks in the van Hove correlation function \cite{kawasaki2013}. We now
 analyze {the} particle hopping rate defined by \cite{lam2018film},
\begin{equation}\label{eq:hop1}
R_1(z)=\frac{P_{\text{hop}}(\Delta t, z) }{\Delta t}.
\end{equation}
Here, $P_{\text{hop}}(\Delta t, z)$ denotes the probability that a particle at layer $z$ hops during a time interval $\Delta t$ and is given by
\begin{equation}  \label{eq:Phop}
P_{\text{hop}}(\Delta t, z) = \langle \Theta(|\mathbf{r}_i(t+\Delta t)-\mathbf{r}_i(t)| -1)\rangle_{t,z},
\end{equation}
where $\Theta(x)=1$ if $x \ge 0$, $\mathbf{r_i}(t)$ is the position of particle $i$ at time $t$ and
the average is taken over time $t$ and all the particles that reside at depth $z$ at the beginning of the time interval. We take a small time interval $\Delta t=10^{-7}$ so that  $R_1(z)$ measures the rate of all particle hops. 

Our main result on particle hops is shown in \Fig{fig:hop1} which displays $R_1(z)$ measured against $z$  from DPLM simulations at a low temperature of $T=0.20$. It is compared with $\tau_\alpha(z)$ reported above and values have been normalized by their approximate bulk values taken at $z=14$. A clear difference between the strengths of the surface effects is seen. Specifically, $\tau_\alpha(z)$ forms a gradient penetrating deep into the interior of the film up to at least $z \simeq 13$. In contrast, a gradient in $R_1(z)$ extends only up to $z \simeq 3$.

\begin{figure}[tb]
 \begin{centering}
  \includegraphics[width=1\columnwidth]{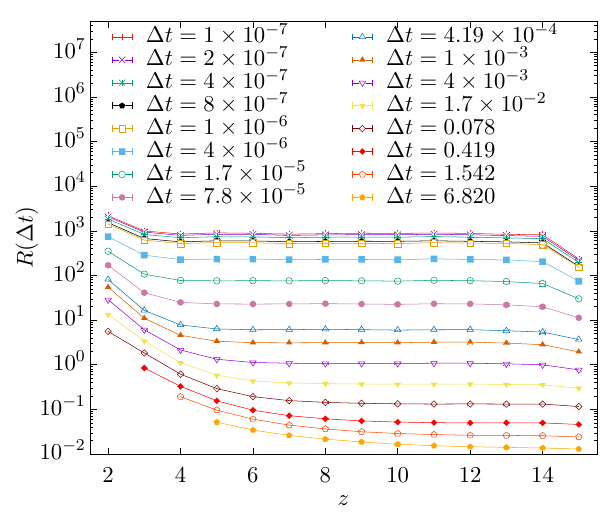}
\caption{Net hopping rates $R(\Delta t)$ against depth $z$ for a film of thickness $h=15$ at different values of $\Delta t$ at $T=0.20$.} 
\label{fig:hop}
\end{centering}
\end{figure}

To justify that time duration $\Delta t=10^{-7}$ used in measuring $R_1(z)$, we next describe more comprehensive results based on other values of $\Delta t$. 
Following a definition applied previously to MD results~\cite{lam2018film}, we  generalize \eq{eq:hop1} to a net hopping rate
$R(z,\Delta t)={P_{\text{hop}}(\Delta t, z) }/{\Delta t}$.
In particular, for the very small time interval $\Delta t=10^{-7}$ used above, the hopping rate $R_1(z)$ is restored. \Fig{fig:hop} plots 
results on $R(z,\Delta t)$ for different $\Delta t$ from DPLM simulations.
Note that we have ensured that $\Delta t$ is small enough by showing only data points for which $P_{\text{hop}}(\Delta t, z) \le 0.5$. 
The results in \fig{fig:hop} qualitatively resemble those from MD simulations \cite{lam2018film}.
For $\Delta t \simeq 10^{-7}$, $R(z,\Delta t) \simeq R_1(z)$ have approximately converged and this signifies that all hops are properly counted.
In contrast, $R(z,\Delta t)$ decreases with $\Delta t$ at larger $\Delta t$.
This is because back-and-forth hops, which are very abundant in the system, then do not contribute to $R(z,\Delta t)$, highlighting  that they also do not contribute to the net dynamics at longer time scales as explained in \cite{lam2018film}.
In particular, the gradient in $R(z, \Delta t)$ penetrates deeper into the bulk for larger $\Delta t$. More precisely, 
$R(z, \Delta t)$ at $\Delta t =6.820$ converges to its bulk value at
$z \agt 12$, in sharp contrast to $R_1(z)$ which converges at a much shallower depth of $z\agt 3$. 
This exemplifies that surface effects for long-time measurements such as $R(z, \Delta t)$ for large $\Delta t$ and 
$\tau_\alpha(z)$ penetrate deeper than short-time values such as $R_1(z)$ \cite{lam2018film}.

\section{Discussions} \label{sec3}

The ECNLE theory based on elastic interactions \cite{schweizer2019} is an insightful analytical approach used in deriving \eq{dex}, providing extensive microscopic details essential for close comparisons with MD simulations and other descriptions of glass relaxation. 
The exponential-of-power-law tail was taken as a unique signature of a long-range elastic field~\cite{ghan2023}.  The fact that it is seen in \fig{fig:de} for the DPLM without elasticity calls into question such a feature as decisive evidence for the elastic picture.

Both the DPLM and the ECNLE theory can account for the exponential-of-power-law tail in the relaxation time, but they essentially produce opposite predictions on the particle hopping rate.
{
  The DPLM results on hopping rates reported above are in good qualitatively agreement with our MD simulations on free-standing polymer films reproted in  \Cite{lam2018film}. In the MD study, we have identified a peculiar layer, referred to as the inner-surface layer, which exhibits a bulk-like particle hopping rate $R_1(z)$ but a surface enhanced mobility as revealed from long-time net hopping rate $R(z,\Delta t)$.
Despite only about three particle diameters thick, its existence including its two apparently contradictory defining features is evident.
The DPLM, as observed from \fig{fig:hop1}, also exhibits such an inner surface layer and is located at $3 \alt z \alt 13$. The thickness is much more pronounced, presumably due to measurement of enhanced mobility based on $\tau_\alpha(z)$ and the thicker film used. Deeper supercooling is expected to increase the thickness of the inner surface layer for both MD and DPLM. 

A natural reason for the bulk-like hopping rate is bulk-like particle hopping energy barriers in the inner surface layer \cite{lam2018film}. 
The close proximity of the inner surface layer to the surface region with a reduced density, which is only four particle diameters away in the MD simulations, implies that particle hops are very localized events depending only on molecular arrangements in the immediate neighborhood.  Thus, barriers dominated by long-range interactions such as elasticity may not be applicable, as argued in  \Cite{lam2018film}. The short-range nature of the interactions in \eq{E} for the DPLM is crucial for the convergence of the hopping rate to the bulk value already $z \simeq 3$ as shown in \fig{fig:hop1}. Note that the dynamics in lattice models with elasticity can be simulated accurately and rather efficiently and particle hopping rates are well-known to depend non-trivially on morphological features and composition at a distance \cite{lam2002}.
}

In contrast, the ECNLE theory  analyzes dynamics dictated by particle activated hopping energy barriers \cite{schweizer2019}. 
{Long-range elasticity introduces a term in the hopping barrier which decays spatially as a power law from the free surface.
  The structural relaxation rate is then follows from the barrier  via a standard Arrhenius relation for activated processes.
This relaxation rate is essentially also the particle hopping rate which must also follow the same Arrhenius relation.}
Therefore, one should predict surface effects on relaxation rate penetrating as deep as those on hopping rates.  It is not clear how the ECNLE theory can be reconciled with MD results in \Cite{lam2018film}.

{
  Back-and-forth motions are known to be important in glassy materials \cite{vollmayr2004,lam2017}.     
  The deeper penetration of surface effects on relaxation time in the inner surface layer has been attributed to a breakdown of such back-and-forth hopping tendency close to the free surface via facilitation \cite{lam2018film}. The DPLM exhibits strongly back-and-forth motions at deep supercooling \cite{zhang2017} and can demonstrate enhanced mobility when those anti-correlations are reduced close to the surface. In contrast, hopping correlation is completely neglected in the ECNLE theory \cite{schweizer2019}. In its present form, it has implemented no mechanism to distinguish between structural relaxation rate and particle hopping rate.

Enhanced mobility has been suggested to propagate into the film via stringlike motions originating from close to the free surface, as revealed by particle trajectory from MD simulations \cite{lam2018film}. Similar facilitation via stringlike motions is also observed from particle displacement profiles in DPLM simulations \cite{zhai2024}.
Assuming that stringlike motions are induced by a fragmented form of voids called quasivoids \cite{yip2020}, we believe that void-induced facilitation accounts for the surface enhanced mobility in glassy films, as illustrated by the DPLM.
}


To conclude, we have measured from DPLM simulations the temperature dependence of four definitions of surface mobile layer thickness in glassy films. Different trends are observed in agreement with previous studies and they follow from different characteristics of the layer-resolved relaxation time. 
We also study the relaxation time including regions deep into the film. A two-component form with a slowly decaying tail closely resembling accurate MD measurements is observed.
Our results show that long-range elastic interactions, which are absent in the DPLM, are not a necessary explanation of the tail. We also demonstrate that the particle hopping rate admits much shallower surface effects in the DPLM, again in agreement with MD simulations. 
{
It has been shown the dynamics is qualitatively similar for the 2D- and 3D-DPLM in bulk geometries~\cite{libo2024}. We thus anticipate that the conclusions drawn from the current results of 2D simulations remain valid in 3D. }

\begin{acknowledgments}
This work was supported by China Postdoc Fund Grant No. 2022M722548, Shaanxi NSF Grant No. 2023-JC-QN-0018, Central University Basic Research Fund Grant No. xzy012023044, {National Natural Science Foundation of China Grant No. 12405042,}   
Hong Kong GRF Grant No. 15303220.
\end{acknowledgments}

\section*{Data Availability Statement}
The data that support the findings of this study are available from the corresponding author upon reasonable request.

\bibliography{glass_short} 

\end{document}